\documentclass[12pt]{article}
 \usepackage{graphicx}
 \usepackage[cp1251]{inputenc}

\begin{document}
 \noindent {\footnotesize\it
   Astronomy Letters, 2022, Vol. 48, No 12, pp. 790--797.}

 \noindent
 \begin{tabular}{llllllllllllllllllllllllllllllllllllllllllllll}
 & & & & & & & & & & & & & & & & & & & & & & & & & & & & & & & & & & & & & &\\\hline\hline
 \end{tabular}

  \vskip 0.5cm
\centerline{\bf\large Link between the Optical and Radio Frames from Gaia DR3 Data}
\centerline{\bf\large and VLBI Measurements}

   \bigskip
   \bigskip
  \centerline {V. V. Bobylev \footnote [1]{e-mail: vbobylev@gaoran.ru}}
   \bigskip

  \centerline{\small\it Pulkovo Astronomical Observatory, Russian Academy of Sciences,}

  \centerline{\small\it Pulkovskoe sh. 65, St. Petersburg, 196140 Russia}
 \bigskip
 \bigskip
 \bigskip

{\bf Abstract}---Based on published data, we have assembled a sample of 126 radio stars with the trigonometric parallaxes and proper motions measured by VLBI and available in the Gaia DR3 catalogue (in fact, Gaia EDR3). Our analysis of the Gaia--VLBI proper motion differences for 84 radio stars based on the model of solid-body mutual rotation has revealed no rotation components differing significantly from zero, $(\omega_x,\omega_y,\omega_z)=(0.06,0.08,-0.10)\pm(0.06,0.07,0.08)$  mas yr$^{-1}.$ Based on the trigonometric parallax differences for 90 stars, we have obtained a new estimate of the systematic offset between the optical and radio frames, $\Delta\pi=-0.022\pm0.017$~mas, and showed that the parallax scale factor is close to unity, $b=1.001\pm0.002$.

 \bigskip
  DOI: 10.1134/S1063773722110032

 \subsection*{INTRODUCTION}
The Gaia trigonometric parallaxes (Prusti et al. 2016) are known to have a slight offset relative to fixed extragalactic sources (quasars). The slight offset that is difficult to take into account remained even in the Gaia DR3 version (Vallenari et al. 2022).

Lindegren et al. (2018) first pointed out the presence of a systematic offset with $\Delta\pi=-0.029$~mas (milliarcseconds) in the Gaia~DR2 parallaxes (Brown et al. 2018) with respect to the inertial reference frame. Later, the presence of such a correction in
the Gaia DR2 parallaxes was confirmed by many authors based on various data with a very good
accuracy. For example, based on a large ($\sim$400 stars) sample of RR Lyrae variables, Muraveva et al. (2018) found a correction $\Delta\pi=-0.057\pm0.006$~mas. Based on 89 detached eclipsing binaries, Stassun and Torres (2018) found a correction
$\Delta\pi=-0.082\pm0.033$~mas. According to these authors, the relative parallax errors for the eclipsing binaries used, on average, do not exceed 5\% and do not depend
on the distance. Riess et al. (2018) estimated $\Delta\pi=-0.046\pm0.013$~mas based on a sample of 50 long-period Cepheids. They used the photometric characteristics of these Cepheids measured onboard the Hubble Space Telescope. Zinn et al. (2019)
found $\Delta\pi=-0.053\pm0.009$~mas by comparing the distances of $\sim$3000 giants from the APOKASC-2 (Apache Point Observatory Kepler Asteroseismology Science Consortium, Pinsonneault et al. 2018) catalogue.

An analysis of the stellar parallaxes from the Gaia EDR3 catalogue (Brown et al. 2021) performed by Lindegren et al. (2021) showed that the parallax zero-point offset depends on the magnitudes and positions of the stars on the celestial sphere. As a result, Lindegren et al. (2021) developed a method of correcting this offset.

Bobylev (2019) compared the stellar parallaxes from the Gaia~DR2 catalogue with those measured by VLBI (very long baseline interferometry). The parameters of the link between the optical and radio frames were derived based on a sample of 88 masers and radio stars. The correction was found to be $\Delta\pi=-0.038\pm0.046$~mas. As can be seen, the result was obtained with a great uncertainty. In this paper it is interesting to repeat the determination of the parallax correction based on a more refined sample of masers
and radio stars to obtain a more reliable estimate (with a smaller error).

A comparison of the absolute proper motions of radio stars from the Gaia catalogue with their VLBI measurements allows the mutual rotation of the optical and radio frames to be controlled. This rotation reflects the quality of the tie-in of both frames to the inertial reference frame (to distant quasars). The absolute proper motions of radio stars from the Gaia~DR2 catalogue were compared with their VLBI measurements, for example, in Bobylev (2019) and Lindegren (2020a, 2020b), where it was concluded that there are no three components of the solid-body mutual rotation vector of these two frames differing significantly from zero.

The goal of this paper is to compare the VLBI parallaxes and proper motions of radio stars with their measurements from the Gaia~DR3 catalogue. It is well known that the stellar parallaxes and proper motions in Gaia~DR3 were simply copied from the Gaia~EDR3 catalogue.

 \section*{DATA}
The VLBI observations of masers associated with young stars and protostars were combined in
the BeSSeL (the Bar and Spiral Structure Legacy Survey \footnote[1]{http://bessel.vlbi-astrometry.org}) project. The project is aimed at determining highly accurate distances to star-forming regions and
studying the structure, kinematics, and dynamics of the Galaxy. The most important contributor here is the American VLBA consisting of ten 25-m dishes with a maximum baseline of more than 8000~km. The observations encompass 6.7 and 12.2 GHz with the emission of methanol (CH$_3$OH) masers and 22.2 GHz with the emission of water (H$_2$O) masers.

Another contributor to the BeSSeL Survey is the European VLBI Network (EVN). Here the longest baselines are $\sim$9000~km, while the 100-m dish at Effelsberg is the largest one in the array. The observations are carried out at frequencies from 6.7 to 22.2 GHz.

In Japan the VLBI observations of masers are performed within the VERA (VLBI Exploration of Radio Astrometry \footnote[2]{http://veraserver.mtk.nao.ac.jp}) program. The interferometer consists of four 20-m dishes distributed over the entire Japan, providing a baseline from 1020 to 2270~km. The observations of H$_2$O masers and, more rarely,
SiO masers are performed at 22.2 GHz and at 43.1 and 42.8 GHz, respectively.

The most important unique property of the VERA dishes is a two-beam receiving system that allows a pair of maser targets and phase reference sources
to be tracked simultaneously. In all other programs
(VLBA, EVN, etc.) the observations of reference
extragalactic objects are carried out at the beginning
and the end of a session by redirecting the dishes,
which then requires additional efforts to take into account the atmospheric distortions. Note that the higher the frequency of observations, the better the astrometric accuracy. Thus, the VLBI observations performed within the VERA program are most accurate compared to the observations within the remaining programs.

The first VLBI parallax measurement for the
source G~339.884-1.259 with the Long Baseline
Array (LBA) interferometer in Australia (Krishnan
et al. 2015) is also known. The interferometer consisted of five dishes with a large diameter (more than 20~m); methanol masers were observed at 6.7~GHz.

Apart from masers, the VLBI observations of radio stars in continuum are of interest. At present, there is complete information approximately for 60 young stars observed within the GOBELINS program (Ortiz-Le\'on et al. 2017) at 5 and 8 GHz---their absolute trigonometric parallaxes and proper motions were measured and their radial velocities are known.

Apart from young stars, the trigonometric parallaxes and proper motions have been determined
by VLBI for dozens of various stars at later evolutionary
stages---giants, red giants, supergiants,
Miras, asymptotic giant branch (AGB) stars, etc.
Such stars are surrounded by extended gas-dust envelopes, where there is maser emission. Their observations are performed at 22 (H2O masers) and 44~GHz (SiO masers).

For the purposes of this paper, the lists of masers
and radio stars from the papers devoted to comparing
the stellar parallaxes and proper motions from
the Gaia~DR2 catalogue with their VLBI measurements
are an important source of data. For example,
Kounkel et al. (2018) compared the parallaxes
using 55 young radio stars located in the Gould Belt.
Bobylev (2019) estimated the parameters of the link
between the optical and radio frames based on a sample
of 88 masers and radio stars. Xu et al. (2019) repeated
Bobylev’s analysis, compiling a list of masers
and radio stars with 108 entries. The list by Xu
et al. (2019) is good in that it gives such important
information about each star as its binarity and specifies
whether it belongs to (i) young objects, (ii) AGB
stars, and (iii) other objects.

Xu et al. (2019) propose not to use AGB stars in the comparison problem, since masers are distributed, occasionally very nonuniformly, in the outer layers of the huge gas–dust envelopes of such stars. This can lead to a non-coincidence of the optical and radio images. One would think that the stars R~Aqr or VY~CMa, for which the Gaia~DR2---VLBI parallax
differences had ``beyond-limit'' values (more than 10~mas), could serve as an example. However, when compared with the Gaia~EDR3 catalogue, the parallax differences for these stars decreased significantly (to less than 2.5 mas) due to the improvement of optical measurements. Xu et al. (2019) also propose not to use binary stars in the comparison problem. We can agree with this.

Note the result of Lindegren (2020a, 2020b), who took 41 bright stars from the list by Xu et al. (2019) with their proper motions from the Gaia~DR2 catalogue having VLBI measurements. Lindegren discarded almost half of these stars by following the criteria developed by him. From the proper motion differences for the remaining 26 stars he estimated the three components of the mutual rotation vector of the optical and radio frames:
$(\omega_x,\omega_y,\omega_z)=(-0.068,-0.051,-0.014)\pm(0.052,0.045,0.066)$~mas yr$^{-1}$, having reached the obvious conclusion about the absence of mutual
rotation components differing significantly from zero. However, of the 26 stars used for the analysis, 5 are AGB stars and 11 are binaries, with two stars (Cyg X-1 and LSI +61 303) being bright components of systems with a black hole.

In this paper we took the list of masers and radio stars from Xu et al. (2019) as a basis, supplementing it with more recent publications devoted to
determining the trigonometric parallaxes and proper
motions of such objects by VLBI. These publications
include Chibueze et al. (2020), Hirota et al. (2020),
Xu et al. (2022), and Sun et al. (2022). Our final list contains 126 entries, with both radio and optical (Gaia EDR3) proper motion measurements being available for 126 stars and the VLBI parallaxes having been measured only for 114 stars.

Note that the list by Xu et al. (2019) contains data on 108 stars, with information only about the VLBI measurements having been available for ten stars. Measurements appeared in the Gaia~DR3 catalogue for some of these stars. As a result, the list of stars with real measuring information increased by 23 stars. The Gaia--VLBI proper motion and parallax
differences for these 23 stars are given in Table 1.

{\begin{table}[t]
\caption[]{\small Data on the 23 stars added to the list by Xu et al. (2019) }
\begin{center} \label{t-data}
\begin{tabular}{|l|r|r|r|c|c|c|}\hline
  Star  & $\Delta \mu_\alpha \cos\delta\pm\sigma$
        & $\Delta\mu_\delta\pm\sigma$\quad\qquad
        & $\Delta \pi\pm\sigma$ \quad\qquad& Ref \\
        & mas yr$^{-1}$~\quad& mas yr$^{-1}$~\qquad&  mas~~\quad\qquad&  \\\hline
DG Tau         &$-1.286\pm0.812$ &$-1.378\pm0.906$ &                 & (1)\\
Parenago 1823  &$-0.064\pm0.235$ &$-1.159\pm0.551$ &                 & (2)\\
Parenago 1844  &$-1.981\pm0.205$ &$ 1.119\pm0.551$ &                 & (2)\\
Parenago 1872  &$-0.168\pm0.130$ &$-0.086\pm0.098$ &                 & (2)\\
Parenago 1896  &$ 0.447\pm0.181$ &$-0.650\pm0.165$ &                 & (2)\\
Parenago 1922  &$-0.912\pm0.207$ &$-0.241\pm0.275$ &                 & (2)\\
V1399 Ori      &$-0.303\pm0.247$ &$-0.280\pm0.522$ &                 & (2)\\
BX Cam         &$-0.082\pm0.212$ &$ 0.828\pm0.451$ &$-0.026\pm0.129$ & (3)\\
OZ Gem         &$ 0.948\pm0.446$ &$-0.832\pm0.349$ &$-0.348\pm0.328$ & (4)\\
R Cnc          &$-0.605\pm0.392$ &$ 0.785\pm0.976$ &$ 0.098\pm0.341$ & (5)\\
X Hya          &$-0.600\pm0.975$ &$ 3.653\pm1.475$ &$ 0.461\pm0.121$ & (5)\\
W Leo          &$ 0.243\pm0.143$ &$ 0.560\pm0.129$ &$-0.152\pm0.110$ & (5)\\
R Hya          &$-0.423\pm1.151$ &$ 4.356\pm1.854$ &$-1.194\pm0.497$ & (5)\\
Y Lib          &$ 0.179\pm2.392$ &$-0.031\pm4.261$ &$-0.023\pm0.097$ & (6)\\
S Ser          &$ 2.868\pm1.427$ &$-1.966\pm2.314$ &$-0.482\pm0.135$ & (5)\\
WR 112         &$ 1.724\pm1.110$ &$ 1.680\pm1.405$ &                 & (7)\\
MAXI J1820+070 &$-0.042\pm0.099$ &$ 0.108\pm0.120$ &$ 0.021\pm0.085$ & (8)\\
V837 Her       &$ 0.372\pm0.788$ &$-0.365\pm0.826$ &$-0.913\pm0.103$ & (9)\\
WR 125         &$-0.821\pm0.500$ &$ 0.597\pm0.600$ &                 & (10)\\
RR Aql         &$ 2.256\pm0.214$ &$ 0.776\pm1.462$ &$-0.497\pm0.139$ & (11)\\
WR 140         &$ 0.755\pm0.202$ &$-0.674\pm0.103$ &                 & (10)\\
WR 146         &$ 2.284\pm0.696$ &$-1.375\pm2.242$ &                 & (10)\\
R Peg          &$ 6.222\pm1.535$ &$-3.797\pm0.924$ &$-0.131\pm0.304$ & (5)\\\hline
\end{tabular}\end{center}
  \small\baselineskip=1.0ex\protect
(1) Rivera et al. (2015); (2) Dzib et al. (2021); (3) Xu et al. (2022); (4) Urago et al. (2020); (5)~Hirota et al. (2020); (6) Chibueze et al. (2019); (7) Yam et al. (2015); (8) Atri et al. (2020); (9)~Chibueze et al. (2020); (10) Dzib and Rodriguez (2009); (11) Sun
et al. (2022).
 \end{table}}
{\begin{table}[t]
\caption[]{\small The components of the mutual rotation vector of the optical (Gaia~EDR3) and radio frames found from the sample of 24 stars from Lindegren (2020a, 2020b) }
\begin{center} \label{t-w1w2w3-Lind}
\begin{tabular}{|l|c|c|c|c|c|c|c|c|c|c|c|c|c|}\hline
 Form of weights $p$ & $N_\star$ & $\omega_x,$~mas yr$^{-1}$ & $\omega_y,$~mas yr$^{-1}$ & $\omega_z,$~mas yr$^{-1}$ \\\hline
 $p=1$      & 24 & $-0.06\pm0.10$ & $-0.08\pm0.12$ & $-0.11\pm0.10$ \\
 $p=1/(\sigma_{\mu_{\rm (Gaia)}}+\sigma_{\mu_{\rm (VLBI)}})$
            & 24 & $+0.07\pm0.06$ & $+0.07\pm0.08$ & $-0.04\pm0.07$ \\
 $p=1/\sqrt{\sigma^2_{\mu_{\rm (Gaia)}}+\sigma^2_{\mu_{\rm (VLBI)}} }$
            & 24 & $+0.07\pm0.06$ & $+0.07\pm0.07$ & $-0.05\pm0.07$ \\
 $p=1/(\sigma^2_{\mu_{\rm (Gaia)}}+\sigma^2_{\mu_{\rm (VLBI)}})$
            & 24 & $+0.17\pm0.04$ & $+0.18\pm0.04$ & $+0.01\pm0.06$ \\ \hline
\end{tabular}\end{center}\small\baselineskip=1.0ex\protect
$N_\star$ --- is the number of stars used..
\end{table}}
{
\begin{table}[t]
\caption[]{\small The components of the mutual rotation vector of the optical (Gaia~EDR3) and radio frames found using 96 stars
 }
\begin{center} \label{t-w1w2w3}
\begin{tabular}{|l|c|c|c|c|c|c|c|c|c|c|c|c|c|}\hline
 Form of weights $p$ & $N_\star$ & $\omega_x,$~mas yr$^{-1}$ & $\omega_y,$~mas yr$^{-1}$ & $\omega_z,$~mas yr$^{-1}$ \\\hline
 $p=1$      & 96 & $-0.10\pm0.09$ & $-0.02\pm0.13$ & $-0.01\pm0.09$ \\
 $p=1/(\sigma_{\mu_{\rm (Gaia)}}+\sigma_{\mu_{\rm (VLBI)}})$
            & 96 & $-0.02\pm0.08$ & $+0.06\pm0.11$ & $-0.04\pm0.07$ \\
 $p=1/\sqrt{\sigma^2_{\mu_{\rm (Gaia)}}+\sigma^2_{\mu_{\rm (VLBI)}} }$
            & 96 & $-0.02\pm0.08$ & $+0.06\pm0.11$ & $-0.04\pm0.07$ \\
 $p=1/(\sigma^2_{\mu_{\rm (Gaia)}}+\sigma^2_{\mu_{\rm (VLBI)}})$
            & 84 & $+0.06\pm0.06$ & $+0.08\pm0.07$ & $-0.10\pm0.08$ \\
 \hline
\end{tabular}\end{center}
\small\baselineskip=1.0ex\protect
$N_\star$ --- is the number of stars left after the rejection by the $3\sigma$ criterion.
\end{table}
}
 \begin{figure}[t]{\begin{center}
  \includegraphics[width=0.7\textwidth]{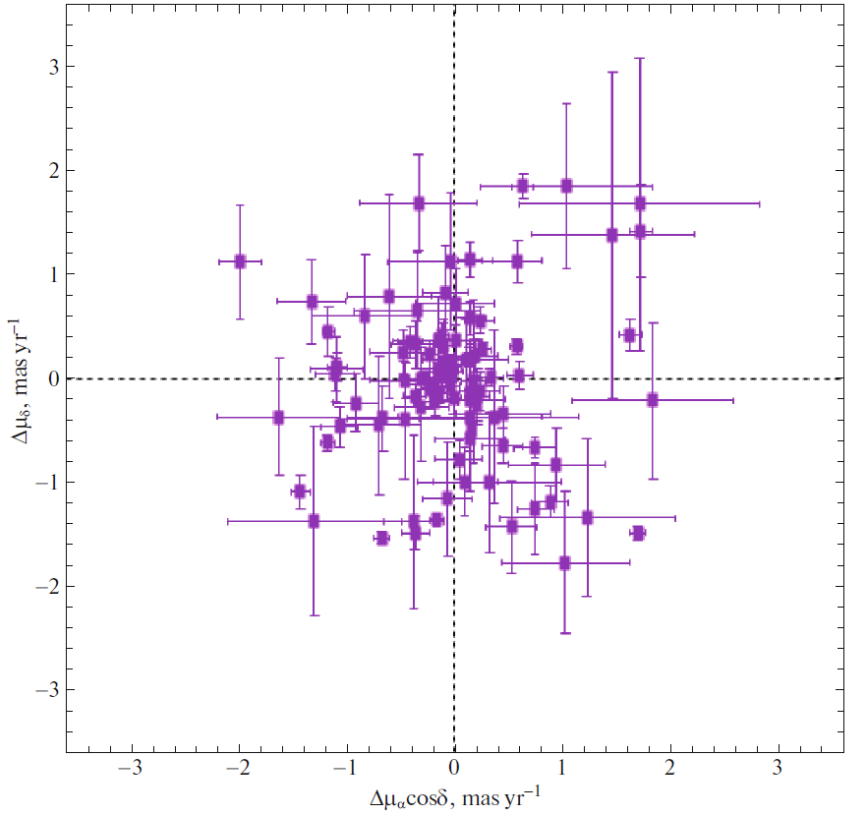}
 \caption{
Gaia--VLBI stellar proper motion differences.
  } \label{f-differ}
\end{center}}\end{figure}
\begin{figure}[t]{\begin{center}
  \includegraphics[width=0.9\textwidth]{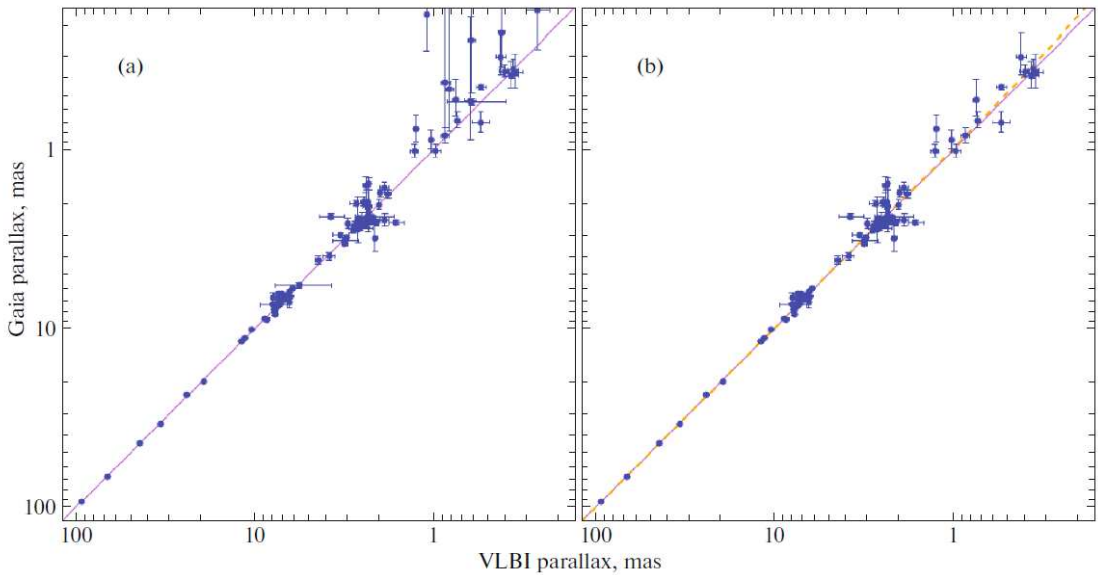}
 \caption{
Parallaxes of radio stars from the Gaia catalogue versus their VLBI parallaxes with relative errors in each case $\sigma_\pi/\pi<100\%$ (a) and errors $\sigma_\pi/\pi<30\%$ (b). The solid line corresponds to a correlation with a coefficient of 1; the dotted line on panel (b) corresponds to $a=-0.022$~mas and $b=1.001$ in Eq. (2).
  } \label{f-par}
\end{center}}\end{figure}
{\begin{table}[t]
\caption[]{\small The parameters of Eq. (2) $a$ and $b$ and the mean $\Delta \pi$ calculated from the Gaia--VLBI differences
 }
\begin{center} \label{t-pi}
\begin{tabular}{|l|c|c|c|c|c|c|}\hline
 Form of solutions & $N_\star$ & $a,$~mas & $b$ & $\Delta \pi,$~mas \\\hline

            &\multicolumn{3}{l}{\qquad$p=1$} & \\
  All             & 100 & $-0.082\pm0.041$ & $1.002\pm0.003$ & $-0.068\pm0.036$ \\
  Without AGB stars   &  84 & $-0.065\pm0.041$ & $1.002\pm0.003$ & $-0.049\pm0.036$ \\
 Only single &  45 & $-0.020\pm0.099$ & $0.983\pm0.021$ & $-0.088\pm0.053$ \\\hline

&\multicolumn{3}{l}{\qquad$p=1/(\sigma_{\mu_{\rm (Gaia)}}+\sigma_{\mu_{\rm (VLBI)}})$} & \\
  All            &  97 & $-0.031\pm0.030$ & $1.002\pm0.003$ & $-0.038\pm0.026$ \\
  Without AGB stars   &  82 & $-0.032\pm0.029$ & $1.001\pm0.003$ & $-0.033\pm0.027$ \\
 Only single &  43 & $-0.106\pm0.071$ & $1.013\pm0.015$ & $-0.047\pm0.040$ \\\hline

&\multicolumn{3}{l}{\qquad$p=1/\sqrt{\sigma^2_{\mu_{\rm (Gaia)}}+\sigma^2_{\mu_{\rm (VLBI)}}}$} & \\
  All            &  97 & $-0.029\pm0.030$ & $1.001\pm0.003$ & $-0.037\pm0.026$ \\
  Without AGB stars   &  82 & $-0.030\pm0.028$ & $1.002\pm0.003$ & $-0.031\pm0.026$ \\
 Only single &  44 & $-0.109\pm0.071$ & $1.014\pm0.015$ & $-0.046\pm0.039$ \\\hline

&\multicolumn{3}{l}{\qquad$p=1/(\sigma^2_{\mu_{\rm (Gaia)}}+\sigma^2_{\mu_{\rm (VLBI)}})$} & \\
  All             &  90 & $-0.020\pm0.022$ & $1.001\pm0.002$ & $-0.022\pm0.017$ \\
  Without AGB stars   &  77 & $-0.015\pm0.021$ & $0.998\pm0.002$ & $-0.019\pm0.018$ \\
 Only single &  41 & $-0.047\pm0.042$ & $1.010\pm0.009$ & $-0.018\pm0.027$ \\
 \hline
\end{tabular}\end{center}
\small\baselineskip=1.0ex\protect
$N_\star$ --- is the number of stars left after the rejection by the $3\sigma$ criterion.
\end{table}
}

 \section*{RESULTS}
 \subsection*{Comparison of the Stellar Proper Motions}
We use the following equations to determine the three angular velocities of mutual rotation of the two frames around the equatorial $\omega_x,\omega_y,\omega_z$ coordinate
axes:
 \begin{equation}
 \begin{array}{lll}
 \Delta\mu_\alpha\cos\delta=
 - \omega_x\cos\alpha\sin\delta - \omega_y\sin\alpha\sin\delta + \omega_z\cos\delta, \\
 \Delta\mu_\delta= + \omega_x\sin\alpha - \omega_y\cos\alpha,
 \label{Gaia-VLBI}
 \end{array}
 \end{equation}
where the Gaia--VLBI differences are on the left-hand sides of the equations. We solve the system of conditional equations (1) by the least-squares method both with unit weights ($p=1$) and with weights inversely proportional to the measurement errors or the squares of the measurement errors. For example, Bobylev (2019) applied weights of the form $p=1/\sqrt{\sigma^2_{\mu_{\rm (Gaia)}}+\sigma^2_{\mu_{\rm (VLBI)}}}$, while Xu et al. (2019)
propose to use weights of the form
 $p=1/(\sigma^2_{\mu_{\rm (Gaia)}}+\sigma^2_{\mu_{\rm (VLBI)}})$.

Initially, based on the system of conditional equations (1), we analyzed the sample of 26 stars selected by Lindegren (2020a, 2020b). In this case, we used the proper motion differences for 24 stars whose absolute values did not exceed 2 mas yr$^{-1}$. The results obtained with the weights of four forms are presented
in Table~2.

Table~3 gives the components of the mutual rotation vector of the optical and radio frames found using 96 stars from our list. The absolute values of the proper motion differences for these stars do not exceed 2 mas yr$^{-1}$. The distribution of differences for these
stars is given in Fig.~1.

 \subsection*{Comparison of the Parallaxes}
We seek the parameters of the link between the parallaxes of the two frames by the least-squares method from the solution of the system of conditional
equations
 \begin{equation}
 \begin{array}{lll}
  \pi_{\rm (Gaia)}=a+b\cdot\pi_{\rm (VLBI)}.
 \label{EQ-1}
 \end{array}
\end{equation}
To compare the trigonometric parallaxes, we used the stars with relative parallax errors no greater than 30\% both in the case of stars from the Gaia~DR3 catalogue and in the case of VLBI measurements.

Table 4 gives the parameters of Eq.~(2) a and b estimated by applying the weights of several forms. The weighted mean Gaia--VLBI parallax difference ${\overline\Delta\pi}$ calculated simultaneously from the same stars is also given. The calculations were carried out by
following the recommendations of Xu et al. (2019) for three cases: (i)~based on the entire sample; (ii)~without using the AGB stars; (iii)~based only on the single stars. It can be seen that the single stars in our sample are too few to obtain a reliable result. The parameters $a$ and ${\overline\Delta\pi}$ characterize one quantity, but ${\overline\Delta\pi}$ is calculated, as can be seen from Table~4, with a smaller error.

Figure 2 plots the parallaxes of radio stars from the Gaia~DR3 catalogue against their VLBI parallaxes. It can be seen from Fig. 2b that both scales are virtually identical at large distances from the Sun. Figure 2a plots the stellar parallaxes from the Gaia catalogue measured with relative errors $\sigma_\pi/\pi<100\%$ against the stellar parallaxes measured
by VLBI with the same constraint, $\sigma_\pi/\pi<100\%$. Figure 2b presents the stellar parallaxes measured with errors $\sigma_\pi/\pi<30\%$. Obviously, to obtain reliable estimates of the sought-for parameters, it is necessary to impose constraints either on the sample radius or on the relative error $\sigma_\pi/\pi$. In this paper we chose the second case.

 \section*{DISCUSSION}
Bobylev (2019) concluded that there are no mutual rotation components between the optical and radio frames differing significantly from zero. In this case, stellar differences no larger than 6 mas yr$^{-1}$ in absolute value were used (Fig.~1 in the paper by Bobylev).
In contrast, in this paper we have a larger number of stars in a more compact region for our analysis, since their absolute values do not exceed 2 mas yr$^{-1}$ (Fig.~1).

The results obtained in this paper using the proper motions of these stars from the Gaia~EDR3 catalogue (Table 2) confirm the conclusion by Lindegren
(2020a, 2020b) that there are no components of the
mutual rotation vector between the frames differing
significantly from zero. It should be noted that the weights
 $p=1/(\sigma^2_{\mu_{\rm (Gaia)}}+\sigma^2_{\mu_{\rm (VLBI)}})$
yield an unreliable result for low statistics.

The results presented in Table 3 were obtained from large statistics, and applying the weights$p=1/(\sigma^2_{\mu_{\rm (Gaia)}}+\sigma^2_{\mu_{\rm (VLBI)}})$ gave the components of the mutual rotation vector with the smallest (compared to
those obtained with other weights) errors.

Note that Bobylev (2019) showed the parallax scale factor $b$ (see relation (2)) to be $1.002\pm0.007$.

Based on a sample of 34 single stars without invoking AGB stars, Xu et al. (2019) estimated the systematic parallax offset for the stars from the Gaia~DR2 catalogue, $\Delta\pi=-0.075\pm0.029$~mas. The estimate of $\Delta\pi=-0.022\pm0.017$~mas obtained in this paper using such weights (Table 4) has a significantly
smaller error. Note that, as practice showed, this
was achieved by applying the following important constraints: (i) on the relative stellar parallax errors ($\sigma_\pi/\pi<30\%$) in both catalogues being compared and (ii) on the absolute value of the parallax differences ($<2$~mas).

A slight parallax zero-point offset with a mean $\Delta\pi\sim-0.019$~mas remained in the Gaia~DR3 version that, in fact, was transferred from the Gaia~EDR3 catalogue. This offset with $\Delta\pi=-0.021$~mas was confirmed by Groenewegen (2021) based on quasars
and with $\Delta\pi=-0.039$~mas when using 75 classical Cepheids. Ren et al. (2021) confirmed this offset with $\Delta\pi=-0.025\pm0.004$~mas by analyzing $\sim$110 000 eclipsing binaries. Based on a sample of
detached eclipsing binaries, which were previously
used by Stassun and Torres (2018) to analyze the
Gaia~DR2 data, these authors found a correction $\Delta\pi=-0.037\pm0.033$~mas in their new paper (Stassun and Torres 2021). Based on a sample of $\sim$300 000
quasars from the Gaia~EDR3 catalogue and using spherical harmonics to analyze the differences, Liao et al. (2021) found a correction $\Delta\pi=-0.021$~mas for a five-parameter solution and $\Delta\pi=-0.027$~mas for a six-parameter solution. A correction $\Delta\pi=-0.028$~mas was found by Wang et al. (2022) when
comparing $\sim$300 000 giants from the LAMOST~DR8 (Large Sky Area Multi-Object Fibre Spectroscopic Telescope) catalogue with the Gaia~EDR3 catalogue.

The dependence of $\Delta\pi$ on the magnitude and coordinates of stars was studied in detail by Lindegren et al. (2021). The correction method proposed
by these authors (for two cases --- five-parameter
and six-parameter solutions) yields good results, as shown by the analysis of various stars (Ren et al. 2021; Huang et al. 2021; Zinn et al. 2021; Wang et al. 2022).

 \section*{CONCLUSIONS}
The list of masers and radio stars from Xu et al. (2019) was supplemented by more recent publications devoted to determining the trigonometric parallaxes
and proper motions of radio stars by VLBI.
The final list includes 126 objects. For 126 stars there
are both radio and optical (Gaia EDR3) proper motion
measurements, while the VLBI parallaxes were measured for 114 stars.

We analyzed the Gaia--VLBI proper motion differences for the radio stars based on the model of solid-body rotation using weights of various forms. The ultimate version of our calculations was obtained with weights of the form
 $p=1/(\sigma^2_{\mu_{\rm (Gaia)}}+\sigma^2_{\mu_{\rm (VLBI)}})$, whose application allowed the sought-for parameters to be estimated with the smallest errors. In particular,
we revealed no mutual rotation components between the optical and radio frames differing significantly from zero,
 $(\omega_x,\omega_y,\omega_z)=(0.06,0.08,-0.10)\pm(0.06,0.07,0.08)$ mas yr$^{-1}.$ This also suggests that each of these frames is excellently tied to the reference
frame of fixed extragalactic sources.

Based on the Gaia--VLBI trigonometric parallax differences for 90 stars, we obtained a new estimate of the systematic offset between the optical and radio
frames, $\Delta\pi=-0.022\pm0.017$~mas, and showed that the parallax scale factor is close to unity, $b=1.001\pm0.002$. Our calculations using the parallax differences
for these stars were performed based on the entire sample and the sample of only single stars. We found no radical differences in the estimates of the sought-for
parameters and, therefore, chose the version with a larger number of stars or, more specifically, using the entire sample.

\bigskip{ACKNOWLEDGMENTS}

I am grateful to the referee for the useful remarks that contributed to an improvement of the paper.

\bigskip\medskip{REFERENCES}{\small

1. P. Atri, J. C. A. Miller-Jones, A. Bahramian, et al.,
Mon. Not. R. Astron. Soc. 493, L81 (2020).

2. V. V. Bobylev, Astron. Lett. 45, 10 (2019).

3. A. G. A. Brown, A. Vallenari, T. Prusti, et al. (Gaia
Collab.), Astron. Astrophys. 616, 1 (2018).

4. A. G. A. Brown, A. Vallenari, T. Prusti, et al. (Gaia
Collab.), Astron. Astrophys. 649, 1 (2021).

5. J. O. Chibueze, R. Urago, T. Omodaka, et al., Publ.
Astron. Soc. Jpn. 71, 92 (2019).

6. J. O. Chibueze, R. Urago, T. Omodaka, et al., Publ.
Astron. Soc. Jpn. 72, 59 (2020).

7. S. A. Dzib and L. F. Rodriguez, Rev. Mex. Astron. Astrofis. 45, 3 (2009).

8. S. A. Dzib, J. Forbrich, M. J. Reid, and K. M. Menten, Astrophys. J. 906, 24 (2021).

9. M. A. T. Groenewegen, Astron. Astrophys. 654, 20 (2021).

10. T. Hirota, T. Nagayama, M. Honma, et al. (VERA
Collab.), Publ. Astron. Soc. Jpn. 70, 51 (2020).

11. Y. Huang, H. Yuan, T. C. Beers, and H. Zhang, Astrophys.
J. 910, 5 (2021).

12. M. Kounkel, K. Covey, G. Suarez, et al., Astron. J. 156, 84 (2018).

13. V. Krishnan, S. P. Ellingsen, M. J. Reid, et al., Astrophys.
J. 805, 129 (2015).

14. S. Liao, Q. Wu, Z.Qi, et al., Publ. Astron. Soc. Pacif. 133, 094501 (2021).

15. L. Lindegren, J. Hernandez, A. Bombrun, et al. (Gaia
Collab.), Astron. Astrophys. 616, 2 (2018).

16. L. Lindegren, Astron. Astrophys. 633, A1 (2020a).

17. L. Lindegren, Astron. Astrophys. 637, C5 (2020b).

18. L. Lindegren, U. Bastian, M. Biermann, et al. (Gaia
Collab.), Astron. Astrophys. 649, 4 (2021).

19. T. Muraveva, H. E. Delgado, G. Clementini, et al., Mon. Not. R. Astron. Soc. 481, 1195 (2018).

20. G. N. Ortiz-Le\'on, L. Loinard, M. A. Kounkel, et al., Astrophys. J. 834, 141 (2017).

21. M. H. Pinsonneault, Y. P. Elsworth, J. Tayar, et al.,
Astrophys. J. Suppl. Ser. 239, 32 (2018).

22. T. Prusti, J. H. J. de Bruijne, A. G. A. Brown, et al.
(Gaia Collab.), Astron. Astrophys. 595, 1 (2016).

23. F. Ren, X. Chen, H. Zhang, et al., Astrophys. J. 911, 20 (2021).

24. A. G. Riess, S. Casertano, W. Yuan, et al., Astrophys. J. 861, 126 (2018).

25. J. L. Rivera, L. Loinard, S. A. Dzib, et al., Astrophys. J. 807, 119 (2015).

26. K. G. Stassun and G. Torres, Astrophys. J. 862, 61 (2018).

27. K. G. Stassun and G. Torres, Astrophys. J. 907, L33 (2021).

28. Y. Sun, B. Zhang, M. J. Reid, et al., Astrophys. J. 931, 74 (2022).

29. R. Urago, R. Yamaguchi, T. Omodaka, et al., Publ. Astron. Soc. Jpn. 72, 57 (2020).

30. A. Vallenari, A. G. A. Brown, T. Prusti, et al. (Gaia Collab.), arXiv: 2208.0021 (2022).

31. C. Wang, H. Yuan, and Y. Huang, Astron. J. 163, 149 (2022).

32. S. Xu, B. Zhang,M. J. Reid, et al.,Astrophys. J. 875, 114 (2019).

33. S. Xu, H. Imai, Y. Yun, et al., arXiv: 2210.02812 (2022).

34. J. O. Yam, S. A. Dzib, L. F. Rodriguez, and V. Rodriguez-G\'omez, Rev. Mex. Astron. Astrofis. 51, 33 (2015).

35. J. C. Zinn, M. H. Pinsonneault, D. Huber, and D. Stello, Astrophys. J. 878, 136 (2019).

36. J. C. Zinn, Astron. J. 161, 214 (2021).
 }
\end{document}